\documentclass[aps,prl,superscriptaddress,twocolumn,showpacs]{revtex4}
\usepackage{amsmath}
\usepackage{graphicx}
\usepackage{amssymb}
\usepackage{bbm, color, soul}

\newcommand*{\tn}[1]{{\textnormal{#1}}}

\usepackage{color}

\begin{document}

\title[]{Double exceptional points generated by the strong imaginary coupling of a non-Hermitian Hamiltonian in an optical microcavity}
\author{Kyu-Won \surname{Park}}
\affiliation{Research Institute of Mathematics, Seoul National University, Seoul 08826, Korea}

\author{Jinuk \surname{Kim}}
\affiliation{Department of Electrical Engineering, Pohang University of Science and Technology, Pohang 37673, Korea}

\author{Kabgyun \surname{Jeong}}
\email{kgjeong6@snu.ac.kr}
\affiliation{Research Institute of Mathematics, Seoul National University, Seoul 08826, Korea}
\affiliation{School of Computational Sciences, Korea Institute for Advanced Study, Seoul 02455, Korea}

\date{\today}

\begin{abstract}
Exceptional points (EPs) have recently attracted considerable attention in the study of non-Hermitian systems and in applications such as sensors and mode switching. In particular, nontrivial topological structures of EPs have been studied intensively in relation to encircling EPs. Thus, EP generation is currently an important issue in several fields. To generate multiple EPs, multiple levels or composite physical systems have been employed with strong real couplings. In this study, we generate multiple EPs on two-level systems in a single microcavity by adopting the strong imaginary coupling of a non-Hermitian Hamiltonian. The topological structures of Riemann surfaces generated by strong imaginary coupling exhibit features that are different from those of Riemann surfaces generated by strong real coupling. The features of these topological structures of Riemann surfaces were verified by encircling multiple EPs and using a Riemann sphere.
\end{abstract}

\maketitle

\section{INTRODUCTION}
Hermitian (closed) quantum systems have real eigenvalues with orthogonal relations. However, real physical systems are non-Hermitian (open), because such systems cannot be isolated from the environment ~\cite{R09}. Openness effects are manifested in the vicinity of a singular point called the exceptional point (EP) in a parameter space, where at least two eigenvalues and their corresponding eigenstates coalesce simultaneously~\cite{W04,T66}.

EPs have recently been studied both theoretically and experimentally in several physical systems such as carbon nanotubes~\cite{GL18}, nano-optomechanical systems~\cite{LP18,WZ21}, photonic crystals~\cite{ZP18,CY21}, optical microcavities~\cite{WS17,CW21}, magnon-polariton systems~\cite{ZL17,HD19}, and electrical circuit resonators~\cite{CH18}.
EPs exhibit several intriguing phenomena associated with parity-time symmetry~\cite{RK18,LT21}, phase transition~\cite{MY16,AR21}, chirality~\cite{TG18,HG22}, Shannon entropy~\cite{KJ18}, and is also useful applications for sensors~\cite{WS17,J14,RJ22}, gyroscopes~\cite{MM18,MA19}, and mode switchings~\cite{DM16,XA22}.

Recently, higher-order EPs~\cite{HA17,IE22} have been studied intensively to enhance the sensitivity of sensors and gyroscopes, because the splitting of eigenvalues against a perturbation ($\epsilon\ll 1$) is proportional to $\epsilon^{1/N}$ in the vicinity of EPs, where $N$ is the order of EPs.
Multiple EPs~\cite{HG16,SA20} are also important, because they can yield more fruitful topological structures than those of a single EP. For example, the encircling of multiple EPs~\cite{JS12,FZ21} or geometric phases around multiple EPs~\cite{SJ12,RD22} have been studied. It should be noted that both higher-order EPs and multiple EPs are generally generated by the strong real coupling of a non-Hermitian Hamiltonian~\cite{JS12,SJ12}. Hence, multiple level (at least three levels)~\cite{JS12,SJ12} or composite physical systems~\cite{XC19,AK22} are required to implement EPs in real systems. Moreover, the topological structures are similar to the functional form $f(z)=(z-z_{0})^{1/N}$, where $N\in\mathbb{N}$ is an integer (typically $N=2,3$), and $z, z_{0}\in \mathbb{C}$.

To the best of our knowledge, no previous study has implemented EPs via strong imaginary coupling of a non-Hermitian Hamiltonian or investigated the relationship between real and imaginary coupling. In this study, we leverage the complex coupling of a non-Hermitian Hamiltonian with a convex combination of the real and imaginary parts in the form of a two-by-two-toy model, and compare it to simulations with the boundary element method for an optical microcavity. Consequently, we implement double EPs on two-level systems in a single microcavity, and verify that previous EPs related to the Landau--Zener avoided crossings~\cite{S08,WA90} are induced by strong real coupling whereas our double EPs related to width bifurcations~\cite{HI13,HI14} are induced by strong imaginary coupling. More interestingly, it was found that our double EPs exhibited distinctive topological structures compared with previous EPs. This distinction fundamentally originates from the different functional forms, i.e., our double EPs are described by $f(z)=\{(z-z_{1})(z-z_{2})\}^{1/2}$, whereas previous EPs are described by $f(z)=(z-z_{0})^{1/2}$.

\section{Introduction to non-Hermitian Hamiltonian}
A closed physical system is described by the Hermitian Hamiltonian, $H_S$. Then, Let us assume that it can interact with its environment and thus becomes an non-Hermitian system (open system). The resulting non-Hermitian system can be described by a non-Hermitian Hamiltonian formulated by
\begin{align}
H_{\rm NH}=H_S+V_{SE} G_E^{(\rightarrow)}V_{ES},
\label{eq1}
\end{align}
where $G_E^{(\rightarrow)}$ is an outgoing Green function in an environment and $V_{SE}$ ($V_{ES}$) is the interaction between the environment (the closed system) and closed system (the environment)~\cite{R09}. Considering two specific interacting eigenstates with negligible interaction with the other states, the non-Hermitian Hamiltonian can be modeled in a two-by-two matrix form as
\begin{align}
H_{\rm NH}=
\begin{pmatrix}
S_{11}+E_{11} & S_{12}+E_{12} \\
S_{21}+E_{21} & S_{2}+E_{22}
\end{pmatrix},
\label{eq2}
\end{align}
where $S_{ij}$ is the matrix representation of $H_{S}$ and $E_{ij}$ is that of $V_{SE}G_E^{(\rightarrow)}V_{ES}$
with respect to the eigenbasis of $H_{S}$. Here, $S_{ij}$ (${i\neq j}$) corresponds to the internal coupling, $E_{ij}$ (${i=j}$) are self-energies (i.e., interaction with the environment itself), and $E_{ij}$ (${i\neq j}$) are called collective Lamb shifts (i.e., interaction via the environment). For convenience, we set $S_{ij}+E_{ij}$ (${i=j}$) to $\xi_{i}\in\mathbb{C}$ and $S_{ij}+E_{ij}$ (${i\neq j}$) to $g\in\mathbb{C}$. Thus, the eigenvalues of the non-Hermitian Hamiltonian are given by:

\begin{align}
\lambda_{\pm}=\frac{\xi_{1}+\xi_{2}}{2}\pm \eta,
\label{eq3}
\end{align}
where $\eta=\sqrt{\frac{(\xi_{1}-\xi_{2})^2}{4}+g^{2}}$. Generally, it is assumed that the coupling constant $g\in \mathbb{C}$ is a Hermitian coupling under the strong real part, i.e., $g_{12}=g^*_{21}$ and ${\rm Re}(g)>{\rm Im}(g)$, and this toy model successfully addresses EPs related to $f(z)=z^{1/n}$-type Riemann surfaces. Here, $N\in\mathbb{N}$ is an integer (typically $N=2, 3$) and $z\in \mathbb{C}$. In this work, however, we extend $g$ to be a non-Hermitian coupling under the strong imaginary part, i.e., $g_{12}\neq g_{21}^{*}$ and ${\rm Im}(g)>{\rm Re}(g)$. This extension supports the new features of a non-Hermitian Hamiltonian and new topological structures of Riemann surfaces.

\section{Proposed physical system}\label{Helm}
\begin{figure}
\centering
\includegraphics[width=8.8cm]{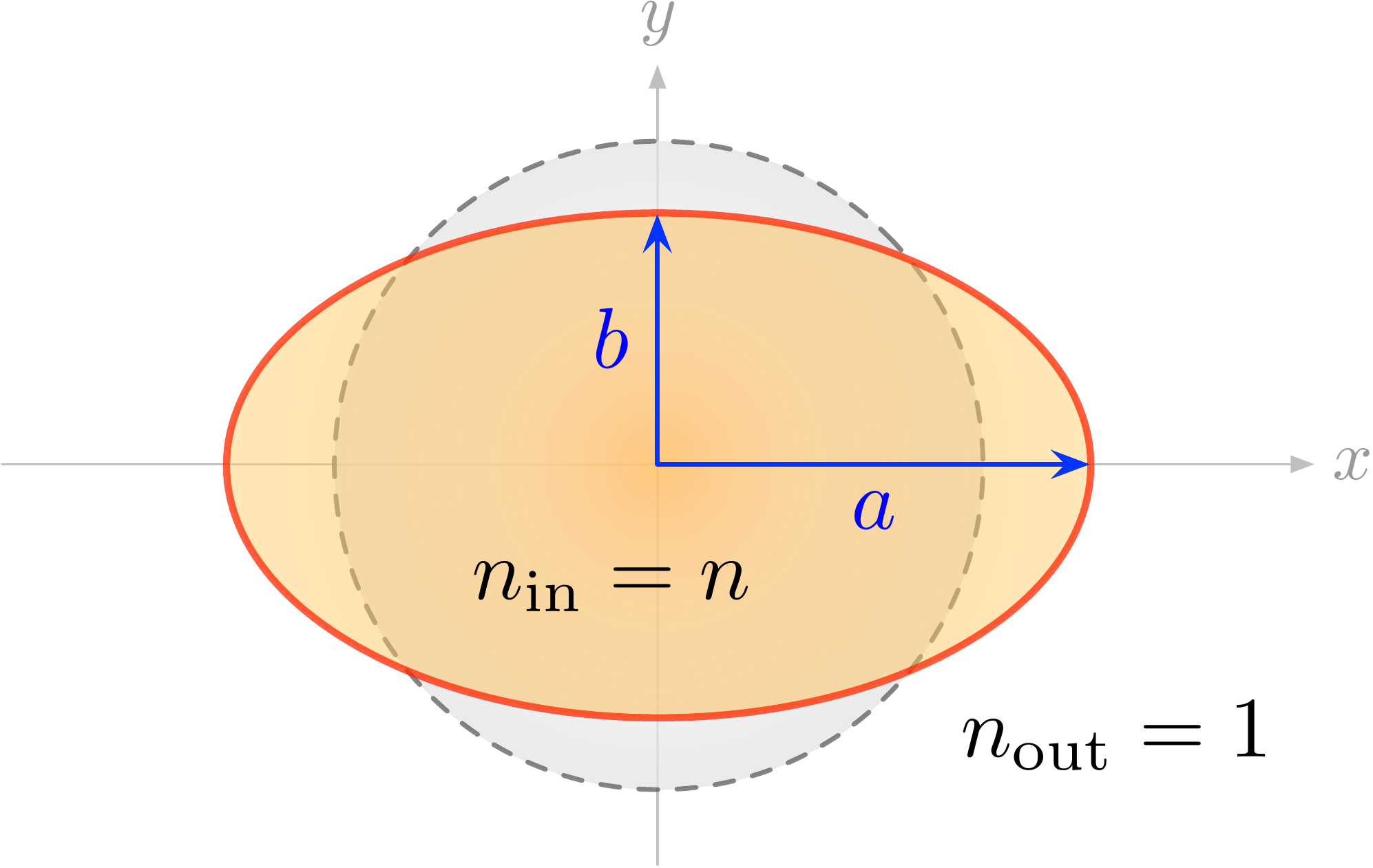}
\caption{A schematic of the boundary of our optical microcavity. The gray-dashed line is a circle ($a=b=1$) deformed to an ellipse with a major axis $a$ and a minor axis $b$. The refractive indices in the inside and outside of the cavity are $n_{\rm in}=n$ and $n_{\rm out}=1$,
respectively.}
\label{Figure-1}
\end{figure}

 Fig.~\ref{Figure-1} shows the geometrical boundary of the optical microcavity. The gray area with the dashed line is a circle, and the pink area with the solid line is the boundary curve of an ellipse with major axis $a$ and minor axis $b$, respectively. The major axis $a$ and minor axis $b$ are defined by $a=1+\chi$ and $b=\frac{1}{1+\chi}$, respectively, as functions of the deformation parameter $\chi$, resulting in a constant area $\pi$. The refractive index on the outside of the microcavity ($n_{\rm out}$) is fixed at $n=1$ (vacuum), whereas the refractive index of the cavity medium ($n_{\rm in}$) can be varied to determine the location of the EPs along with the deformation parameter $\chi$ in the parameter plane. The eigenvalues and their eigenfunctions are obtained by solving the Helmholtz equation $\nabla^{2}\psi+n^{2}k^{2}\psi=0$, using the boundary element method (BEM)~\cite{W03} for the transverse magnetic modes in the elliptical two-dimensional cavity in the $x$ and $y$ planes. Here, $k$ is the wave number and $\psi$ is the $z$-component of the electric field.

In this work, the elliptical optical two-dimensional microcavity is considered as an open system, which can be described by a non-Hermitian Hamiltonian. This system is appropriate for studying non-Hermitian effects, because the closed elliptical cavity is an integrable system without any avoided crossings, resulting in Poisson distributions~\cite{H99,F10}.

Accordingly, including the internal coupling $S_{ij}$ (${i\neq j}$) is not required, and the off-diagonal elements in Eq.~(\ref{eq2}) are just $E_{ij}$ (${i\neq j}$), i.e., we only consider the avoided crossing from the effects of openness.

\begin{figure*}
\centering
\includegraphics[width=16.0cm]{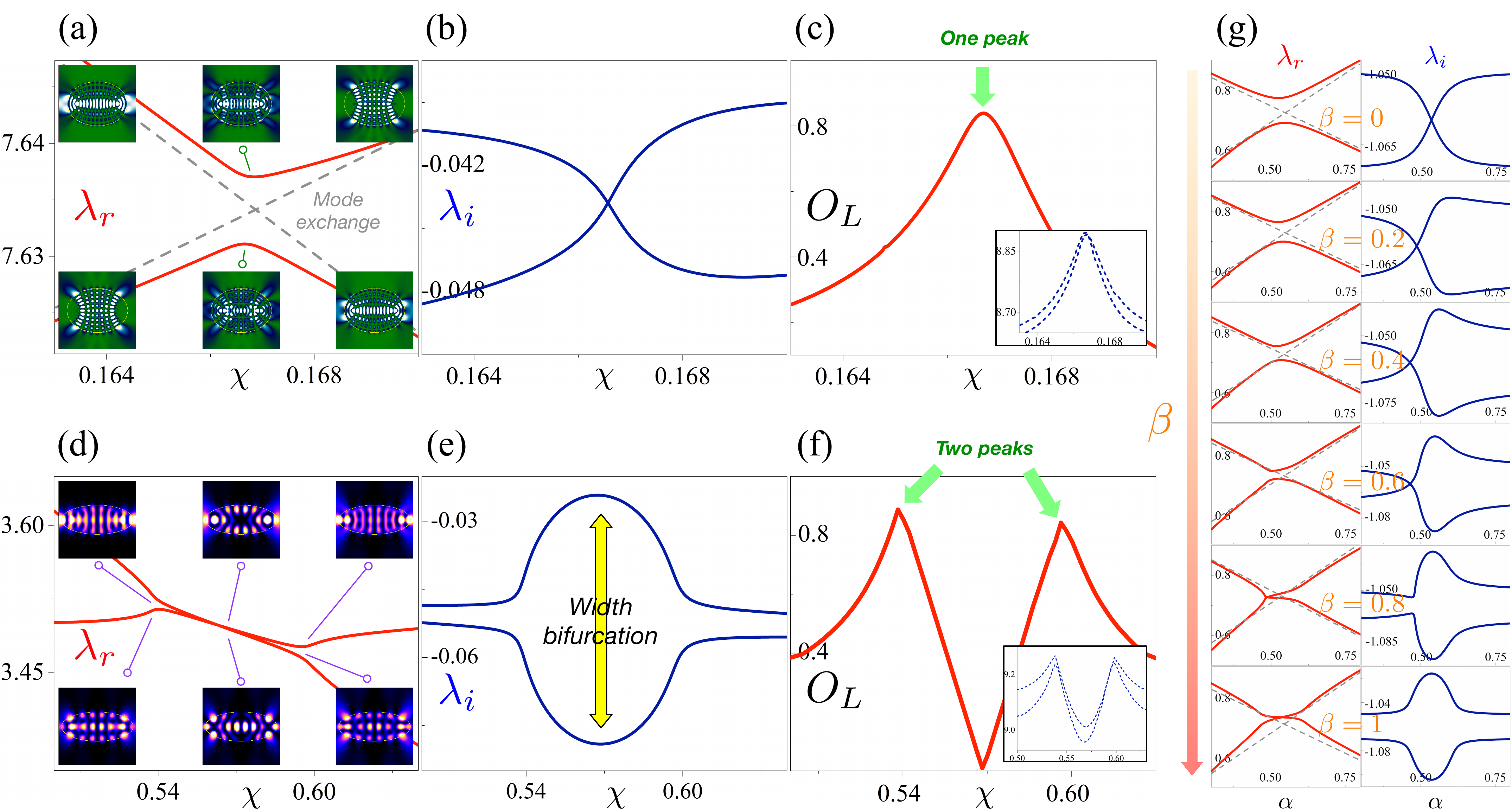}
\caption {Typical Landau--Zener avoided crossing (upper panels) vs. width bifurcation (lower panels). The insets are intensity plots of eigenstates corresponding to the eigenvalues. (a) Real parts of eigenvalues show avoided crossing with exchange of eigenstates. (b) Imaginary parts of eigenvalues show mode crossing. (c) An overlap integral of two interacting eigenstates is maximized around the center of the avoided crossing. The inset (Shannon entopy) exhibits similar behavior to the overlap integral. (d) Real parts of eigenvalues show crossing. (e) Imaginary parts of eigenvalues are bifurcated around two critical points. (e) An overlap integral of two interacting eigenstates is maximized around two edge of the bifurcations. The inset (Shannon entopy) exhibits similar behavior to the overlap integral. (g) Transition from Landau--Zener avoided crossing to width bifurcation is displayed as function of $\beta$.}
\label{Figure-2}
\end{figure*}

\section{Extension from Hermitian coupling to non-Hermitian coupling in a toy model}
In this section, the coupling $g$, which is typically defined as Hermitian coupling, is extended to non-Hermitian coupling in the form of a $2\times2$ non-Hermitian Hamiltonian toy model. To perform the calculations of the toy model, following Ref.~\cite{HI14}, we set the real parts of ${\xi}$ (i.e., $\xi^{r}$) as a function of $\alpha$ as $\xi_{1}^{r}(\alpha)=1-\frac{\alpha}{2}$ and $\xi_{2}^{r}(\alpha)=\sqrt{\alpha}$, while the imaginary part of $\xi$ (i.e., $\xi^{i}$) is fixed in the considered range of parameters. Moreover, the non-Hermitian coupling term $g$ is exponentially sensitive to the relative difference between the two values ($\xi_{1},\xi_{2}$), yielding the following form: $g(\alpha;\xi_{1}, \xi_{2})=g_{c}\Lambda_{\alpha}(\xi_{1}, \xi_{2})$. Here, $g_{c}$ is the coupling coefficient and the sensitivity $\Lambda_{\alpha}(\xi_{1}, \xi_{2})$ is defined as $\Lambda_{\alpha}(\xi_{1}, \xi_{2})=\exp\left[-\{\xi_{1}(\alpha)-\xi_{2}(\alpha)\}^2\right]$~\cite{HI14}. In general, the coupling coefficient $g_{c}$ is complex.

\subsection{Comparison between Landau--Zener interaction and width bifurcation}
To validate the effects of extension from a Hermitian coupling to a non-Hermitian one, especially to verify the role of the ratio between the imaginary part and real part, we examine the typical Landau-Zener avoided crossing and less known width bifurcation simultaneously. Upper panels (a), (b), and (c) in Fig. ~\ref{Figure-2} correspond to Landau-Zener avoided crossing~\cite{JE29,KS18}, whereas the lower panels (d), (e), and (f) in Fig.~\ref{Figure-2} correspond to width bifurcation~\cite{HI14,HI13}. Around $\chi\simeq 0.167$, the real parts of eigenvalues, $\lambda_{r}$, in Fig.~\ref{Figure-2}(a) show an avoided crossing with an exchange of eigenstates, whereas the imaginary parts of the eigenvalues, $\lambda_{i}$, in Fig. ~\ref{Figure-2}(a) show the crossing. The insets in Fig.~\ref{Figure-2}(a) show the intensity plots of the eigenfunctions corresponding to each eigenvalue. Furthermore, to study the relation with exceptional points, we consider the overlap integrals of two interacting eigenstates and the Shannon entropies of the eigenstates. The overlap integral is defined as follows:

\begin{align}
O_{L}(i,j)= \frac{1}{X_{i}X_{j}}\int\int dxdy |\psi^{*}_{i}(x,y)\psi_{j}(x,y)|,
\label{eq4}
\end{align}
where $X_{i}=\sqrt{\int dxdy|\psi_{i}(x,y)|^{2}}$ denotes the normalization factor.
The results are presented in Fig. ~\ref{Figure-2}(c), and its value was maximized around the center of the avoided crossing ($\chi\simeq 0.167$). The Shannon entropy of the normalized intensity of the eigenfunctions is defined by $H(\rho)=-\int\rho(r)\log\rho(r)dr$ and resulting $H(\rho)$ are shown in the inset of Fig. ~\ref{Figure-2}(c)~\cite{KJ18}, which exhibits a similar behavior to that of $O_{L}$. These behaviors directly reveal that the EP is located around $\chi\simeq 0.167$.

Furthermore, the width bifurcation (lower panels) shows significantly different behavior compared with the upper panels. First, the real parts of the complex eigenvalues ($\lambda_{r}$) are crossed, and interestingly, the real parts ($\lambda_{r}$) stick together in the finite parameter regions. This behavior is quite different from Landau-Zener interactions, which exhibit avoided crossing in some narrow regions. The insets in Fig.~\ref{Figure-2}(d) show the intensity plots of the eigenfunctions corresponding to each eigenvalue. The imaginary parts of the complex eigenvalues ($\lambda_{i}$) reveal more intriguing phenomena, i.e., $\lambda_{i}$'s are bifurcated around two critical points, and the relative difference between these two $\lambda_{i}$'s is maximized at the center of the two critical points. The dramatically increasing (decreasing) decay is known as super-radiance (sub-radiance) in quantum optics~\cite{HI14,IJ15}, and is related to the exceptional ring of the Dirac cone~\cite{B15}.
In Fig.~\ref{Figure-2}(f), the overlap integral of the two interacting eigenstates $O_{L}$ is maximized around the two edges of the bifurcations. The inset (Shannon entropies) exhibits a behavior similar to $O_{L}$. As mentioned previously, these two peak values indicate that there are two EPs around these two peaks.

According to these observations, the upper panels associated with the Landau--Zener interaction and the lower panels associated with the width bifurcation exhibit distinctive features. Accordingly, we infer that these phenomena originate from different physical influences, and that they cannot be related to each other. However, an extension from Hermitian coupling to non-Hermitian coupling can resolve these problems. For this approach, we can constitute a complex coefficient ($g_{c}$) under a convex combination of real and imaginary parts, resulting in the following form:
\begin{align}
g(\alpha,\beta)=g_{c}\left[(1-\beta)+\iota\beta\right]\Lambda_{\alpha}(\xi_{1}, \xi_{2}),
\label{eq5}
\end{align}
where $g\in \mathbb{C}$, $g_{c}$, $\beta \in \mathbb{R}$, $\iota=\sqrt{-1}$, and $0\leq \beta \leq1$.

By employing $g(\alpha,\beta)$, we calculate the eigenvalues ($\lambda_{\pm}$) with the initial values of $g_{c}=0.043$, $\xi_{1}^{i}=1.05$, $\xi_{2}^{i}=1.07$, and the results are shown in Fig. ~\ref{Figure-2}(g)). The results show a smooth transition from the Landau--Zener interaction to width bifurcation as a function of increasing $\beta$. The coupling coefficient $g_{c}$ is a pure real value at $\beta=0$, and it corresponds to Landau--Zener interactions, as shown in Fig. ~\ref{Figure-2}(a) and (b)). At $\beta=0.8$, $\lambda_{r}$ s are crossed, and $\lambda_{i}$ s clearly exhibit a width bifurcation. When $\beta$ reaches a value $1$, the coupling $g_{c}$ becomes a pure imaginary value, which corresponds to the width bifurcation, as shown in Fig. ~\ref{Figure-2}(d) and (e).

 In this way, the Landau-Zener interaction can be smoothly transformed to width bifurcation by exploiting the non-Hermitian coupling coefficient under a convex combination of the real and imaginary parts.

\subsection{Double exceptional points in a non-Hermitian Hamiltonian toy model}
We verified that there were two similar peaks in Fig. ~\ref{Figure-2}(f), as well as in its inset, and it is assumed that these indicate two EPs around the two peaks. A simple non-Hermitian Hamiltonian toy model with a non-Hermitian coupling $g(\alpha,\beta)$ can verify this conjecture. The definition of EPs indicates that they can be found when $\eta=0$. Hence, to find double EPs for our toy model easily, we set $\xi_{1}^{i}=\xi_{2}^{i}=1.05$ and $g_{c}=0.05$, and solve the simple algebraic equation:
\begin{align}
\eta(\alpha,\beta)&=0=\sqrt{\frac{(\xi_{1}-\xi_{2})^2}{4}+g^{2}}\\ \nonumber
&={\tilde{\alpha}}^{2}+\frac{1}{100}\left[2\beta-1+2\imath\beta(1-\beta)\right]e^{-2\tilde{\alpha}^2},
\end{align}
where $\tilde{\alpha}=1-\frac{\alpha}{2}-\sqrt{\alpha}$.

\begin{figure}
\centering
\includegraphics[width=8.8cm]{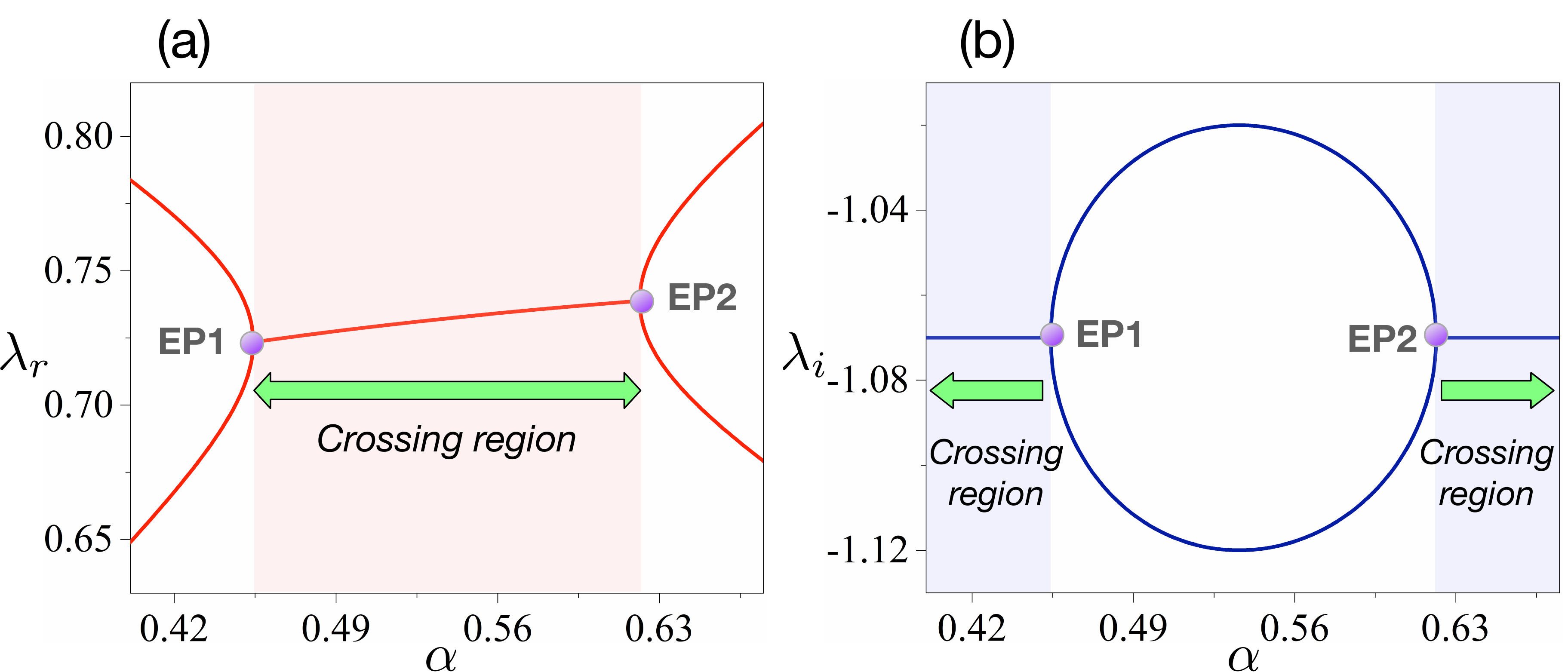}
\caption {Eigenvalue trajectories associated with the double EPs of the toy model of a non-Hermitian Hamiltonian with a non-Hermitian coupling under strong imaginary part. (a) Real parts of complex eigenvalues. (b) Imaginary parts of complex eigenvalues. The double EPs are located at $\alpha=0.454$ and $ \alpha=0.621$, respectively.}
\label{Figure-3}
\end{figure}

A simple analytical calculation yields the double EPs {\bf EP1} ($\alpha=0.454, \beta=1$) and {\bf EP2} $(\alpha=0.621, \beta=1$). This result is confirmed when the eigenvalue trajectories are traced as a function of $\alpha$ with a fixed $\beta=0$.
Fig.~\ref{Figure-3} shows the double EPs located at $\alpha=0.454$ and $ \alpha=0.621$. Two eigenvalue trajectories of $\lambda_{r}$ are crossed (with the same values) throughout the parameter range between the two EPs in Fig. ~\ref{Figure-3}(a). In this range, the split term $\eta$ is a purely imaginary value, because the coupling term $g$ has negative values under the condition $g>\frac{\xi_{1}-\xi_{2}}{2}$. Thus, $\eta$ induces width bifurcation in that region. By contrast, the two eigenvalue trajectories of $\lambda_{i}$ in Fig. ~\ref{Figure-3}(b) are complementary to that for $\lambda_{r}$ in Fig. ~\ref{Figure-3}(a), such that they cross beyond the double EPs. Consequently, it appears that there are two distinct crossing regions for the two eigenvalue trajectories of $\lambda_{i}$, in contrast to Fig. ~\ref{Figure-3}(a). This discrepancy is resolved in Sec. VI.

\section{Successive transitions of avoided crossings in a microcavity and toy model}
\begin{figure}
\centering
\includegraphics[width=8.8cm]{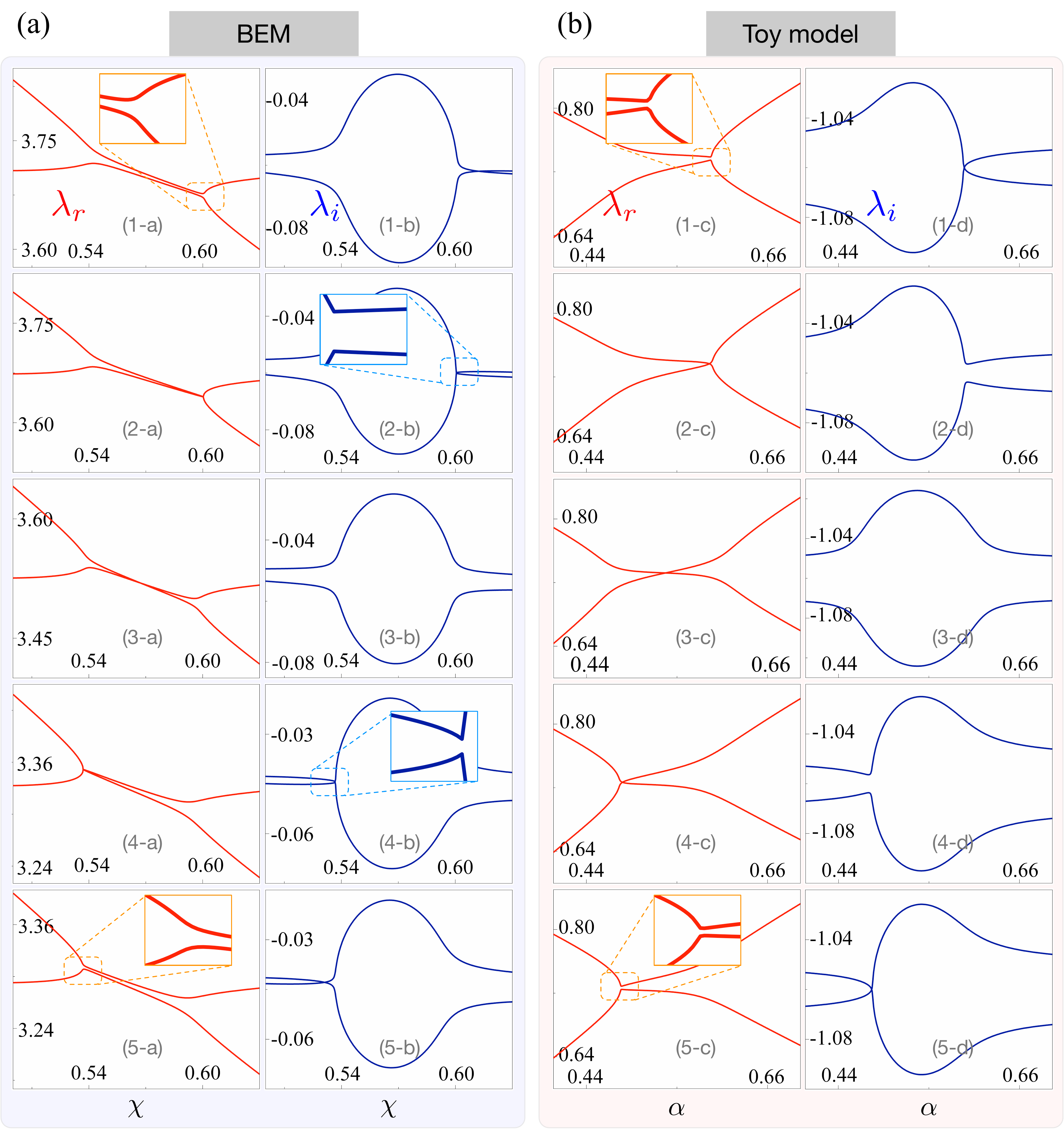}
\caption {Some representative eigenvalue trajectories according to Table I. Left panels are eigenvalue trajectories obtained from our BEM simulation. Right panels are eigenvalue trajectories obtained from the toy model Hamiltonian. Both panels show similar behavior.}
\label{Figure-4}
\end{figure}

\begin{table*}
\center
\begin{tabular}{c|c|c|c|c|c|c|c}
\hline\hline
{} & $n_{\tn{in}}$ & $g_{c}$ & $\beta$ & $\gamma_1$ & $\gamma_2$ & $\lambda_r$ & $\lambda_i$ \\ \hline
Class 1 & 2.6 & 0.043 & 0.76 & $1.05$ & $1.07$ & A.C. & C. \\ \hline
Class 2 & 2.6257 & 0.043 & 0.78 & $1.05$ & $1.07$ & C. & A.C. \\ \hline
Class 3 & 2.74 & 0.043 & 1 & $1.05$ ($1.07$) & $1.07$ ($1.05$) & C. & A.C. \\ \hline
Class 4 & 2.9036 & 0.043 & 0.78 & $1.07$ & $1.05$ & C. & A.C. \\ \hline
Class 5 & 2.94 & 0.043 & 0.76 & $1.07$ & $1.05$ & A.C. & C. \\
\hline\hline
\end{tabular}
\bigskip
\caption{Some representative parameters for successive transitions of the avoided crossings in the microcavity and toy model. Class $1$ corresponds to the avoided crossing of $\lambda_{r}$ with crossing of $\lambda_{i}$. Class 2, 3 and 4 correspond to the avoided crossing of $\lambda_{i}$ with crossing of $\lambda_{r}$. Class $5$ corresponds to the avoided crossing of $\lambda_{r}$ with crossing of $\lambda_{i}$.}
\label{table-1}
\end{table*}

We identified double EPs analytically in the non-Hermitian Hamiltonian toy model described in the previous section for a simple case. However, the analytical solutions of the non-Hermitian Hamiltonian matrix elements are not explicitly known in microcavities, which are common in conventional open physical systems. Hence, we rely on numerical methods to specify the locations of the EPs. A good strategy can be coarse scanning a parameter plane until the transition of the avoided crossing occurs, because the EP is a singular point where the transition between strong and weak interactions occurs ~\cite{S08,WA90}. Thus, an avoided crossing of the real part with the crossing of the imaginary part is converted into an avoided crossing of the imaginary part with the crossing of the real part when going across the EPs and vice versa.

\begin{figure*}
\centering
\includegraphics[width=\textwidth]{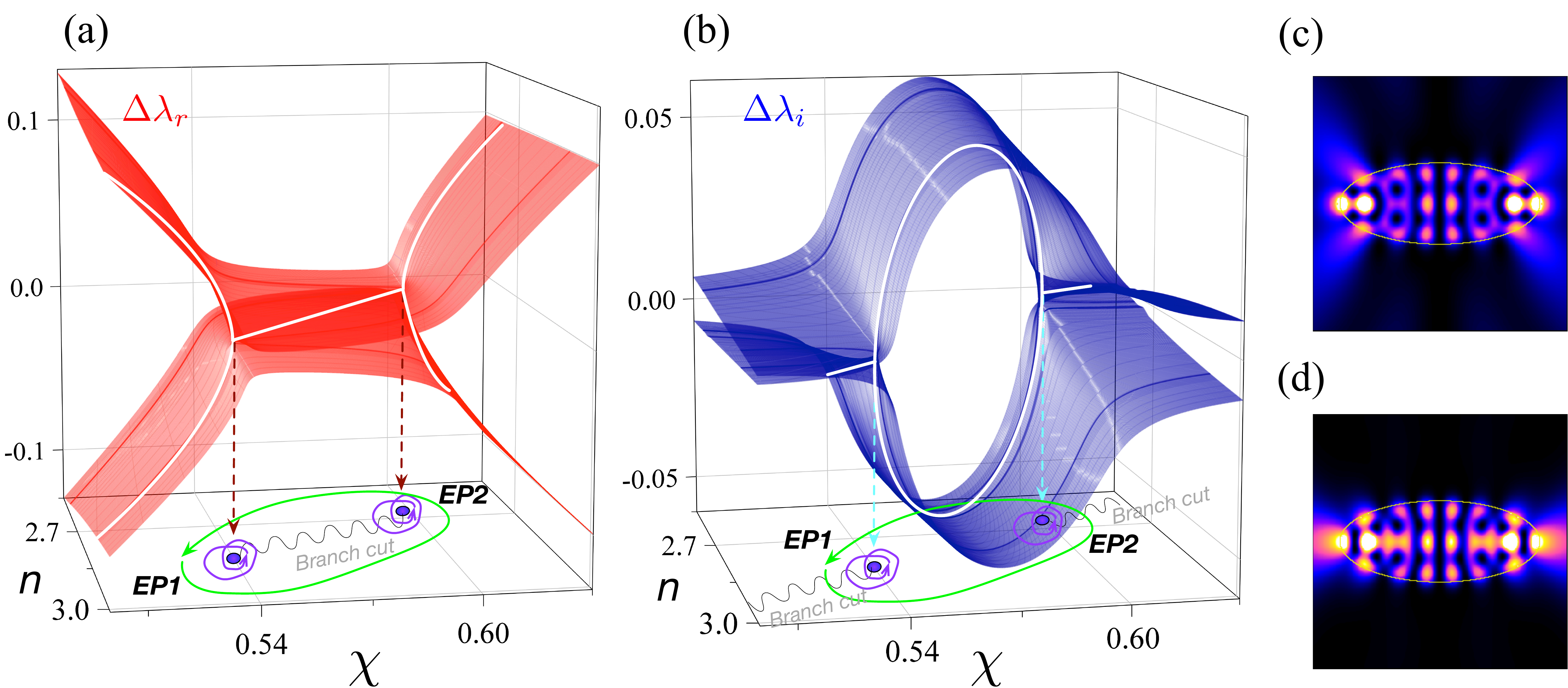}
\caption {A new topological structure $f(z)=\sqrt{(z-z_{1})(z-z_{2})}$ of the double EPs in an optical microcavity are displayed in a parameter plane. (a) Riemann surfaces of real parts of complex eigenvalues. (b) Riemann surfaces of imaginary parts of complex eigenvalues. The double EPs are located at $n\simeq 2.6257, \chi=0.601$, and $n\simeq 2.9036, \chi=0.53$, respectively. Encircling each EP (including the double EPs) are schematically illustrated by purple arrows (green arrows) on the bottom of (a) and (b). The figures (c) and (d) are intensity plots of eigenfunctions at {\bf EP1} and {\bf EP2}, respectively.}
\label{Figure-5}
\end{figure*}

Consequently, we found five representative classes in both the BEM and toy models, as shown in Fig.~\ref{Figure-4}. The left panels are the eigenvalue trajectories obtained from our BEM simulation, whereas the right panels are the eigenvalue trajectories obtained from the toy model Hamiltonian for some specific parameters (shown in Table I). For the BEM, we obtained the eigenvalues by scanning the deformation parameter $\chi$ in the range of $0.05\leq\chi \leq 0.63$ at each fixed $n_{\rm in}$. Note that these two panels show similar behavior in each class. Five representative classes were classified to capture the transition of avoided crossing.
Class $1$ corresponds to the avoided crossing of the real part: (1-a) and (1-c), and the crossing of the imaginary part: (1-b) and (1-d). On the contrary, Classes 2,3, and 4 correspond to the avoided crossing of the imaginary part: (2-b, 3-b, 4-b) and (2-d, 3-d, 4-d), with the crossing of the real part: (2-a, 3-a, 4-a) and (2-c, 3-c, 4-c). Finally, Class 5 again corresponds to the avoided crossing of the real part: (5-a) and (5-c), with the crossing of the imaginary part: (5-b), (5-d).
Importantly, for the left panels (BEM), a careful examination reveals that the transitions of avoided crossing occur around two critical points at $\chi\simeq 0.6$ and $\chi\simeq 0.53$, respectively. These two critical points correspond to the double EPs (as shown in Fig.~\ref{Figure-5}).

For a deeper understanding of these phenomena, we present some representative parameters for the successive transition of avoided crossing in both the microcavity and toy models. Since the refractive index $n_{\rm in}$ only increases as the class number increases, we can hardly obtain any meaningful information related to the transition of the avoided crossing in the BEM simulations. However, various parameters of the toy model can provide useful physical information. First, these types of transitions can occur when the imaginary part of the complex coefficient of coupling is larger than the real part of the complex coefficient of coupling throughout the process. This condition is essential for maintaining the structure of the width bifurcation, as shown in Fig. ~\ref{Figure-1}(g)). It should be noted that the key to successive transitions of the avoided crossings is related to the crossover of the critical value of $\beta_{c}$ in the range $0.76\leq\beta\leq 0.78$. In our examples, the value of $\beta_{c}$ was approximately $0.765$. Therefore, Class 1 has $\beta$ less than $\beta_{c}$. The value of $\beta$ is larger than $\beta_{c}$ in Class 2, and increases until $\beta=1$ in Class 3. Then, $\beta$ decreases towards $\beta_{c}$ in Class 4. In Class 5, it becomes smaller than $\beta_{c}$. Moreover, the relative magnitude of $\gamma_{i}$ determines the crossover position. When $\gamma_{2}>\gamma_{1}$, crossover occurs at approximately $\alpha=0.61$, whereas for $\gamma_{1}>\gamma_{2}$, crossover occurs approximately $\alpha=0.52$.

\section{Riemann surfaces for double EPs in an optical microcavity}
\subsection{Topological structure of Riemann surfaces for double EPs}
We analyzed the successive transitions associated with the five representative avoided crossings in the range of $0.5\leq \chi \leq 0.63$ and $2.6\leq n \leq 2.94$, and verified the two critical points around $n\simeq 2.6257, \chi=0.601$ and $n\simeq 2.9036, \chi=0.53$. Accordingly, by obtaining the whole eigenvalue trajectories by scanning the parameter plane that includes this range, we obtain new topological structures of the double EPs. To display the topological structured of double EPs more clearly, we consider the relative differences $\Delta\lambda_{\pm}$ from the eigenvalues $\lambda_{\pm}$ with respect to their mean values $\lambda_{\rm AV}=\frac{\lambda_{+}+\lambda_{-}}{2}$ instead of the eigenvalue $\lambda_{\pm}$: $\Delta\lambda_{\pm}=\lambda_{\pm}-\lambda_{\rm AV}$. In this way, the Riemann surfaces of the real parts are displayed in Fig.~\ref{Figure-5}(a), whereas those of the imaginary parts are shown in Fig.~\ref{Figure-5}(b). Double EPs ({\bf EP1} and {\bf EP2}) are located at $(n\simeq 2.6257 ,\chi=0.6001)$ and $(n\simeq 2.9036 ,\chi=0.5372)$, respectively. Moreover, the branch cut of the Riemann surfaces for the real parts is a curve directly joining these two EPs, whereas that for the imaginary part is the two curves in the outward direction to the branch cut of the Riemann surfaces for real parts. Accordingly, we immediately notice that their topological structures are manifestly different from that of typical Riemann surfaces, such as $f(z)=(z-z_{0})^{1/N}$. It can be argued that the Riemann surfaces shown in Fig.~\ref{Figure-5} have topological structures equivalent to the following function:
\begin{align}
 f(z)=\sqrt{(z-z_{1})(z-z_{2})}.
 \label{eq7}
\end{align}

First, the derivative of $f(z)$ with respect to $z$ diverges at $z=z_{1}$ and $z=z_{2}$. This guarantees the existence of two branch points (two EPs) at $z=z_{1}$ and $z=z_{2}$. Furthermore, the encircling EPs, which are schematically illustrated at the bottom of Fig. ~\ref{Figure-5} validate this argument. Two cyclic variations around each EP are required for $\lambda_{r}$ and $\lambda_{i}$ to return to their original values on the eigenvalue surfaces, as indicated by the purple arrows. Moreover, a single cyclic variation is required to return to the original values when considering the loop including both branch points (double EPs), as indicated by the green arrows. These are essential topological features of complex-valued functions like Eq.~(\ref{eq7}). Note that both Fig.~\ref{Figure-5}(a) and 5(b) show these features simultaneously. However, it should be noted that even though Fig.~\ref{Figure-5}(a) and 5(b) have different topological structures for each branch cut, they result in these features equivalently.

When moving around the loop, including the double EPs (green arrows), Fig.~\ref{Figure-5}(a) and 5(b) reveal different behaviors to each other. In other words, for the Riemann surfaces shown in Figs. ~\ref{Figure-5}(a), the cyclic variations do not cross any branch cuts throughout the procedure, resulting in them staying on the same Riemann sheet. By contrast, for the Riemann surfaces shown in Fig. ~\ref{Figure-5}(b), the cyclic variations go across branch cuts twice throughout the procedure, i.e., when crossing a branch cut from a Riemann sheet, it lands on the other Riemann sheet and then returns to the original Riemann sheet after crossing the other branch cut. The mathematical structure for describing these double EPs and their branch cuts is presented in the next subsection.

It is worth mentioning that the similarities between Fig.~\ref{Figure-3} and Fig. ~\ref{Figure-5} can be confirmed when the Riemann surfaces are cut along a curve containing branch cuts in the parameter space, as indicated by the white solid lines.

\subsection{Riemann sphere for understanding of topological structure of double EPs}
\begin{figure}
\centering
\includegraphics [width=8.5cm] {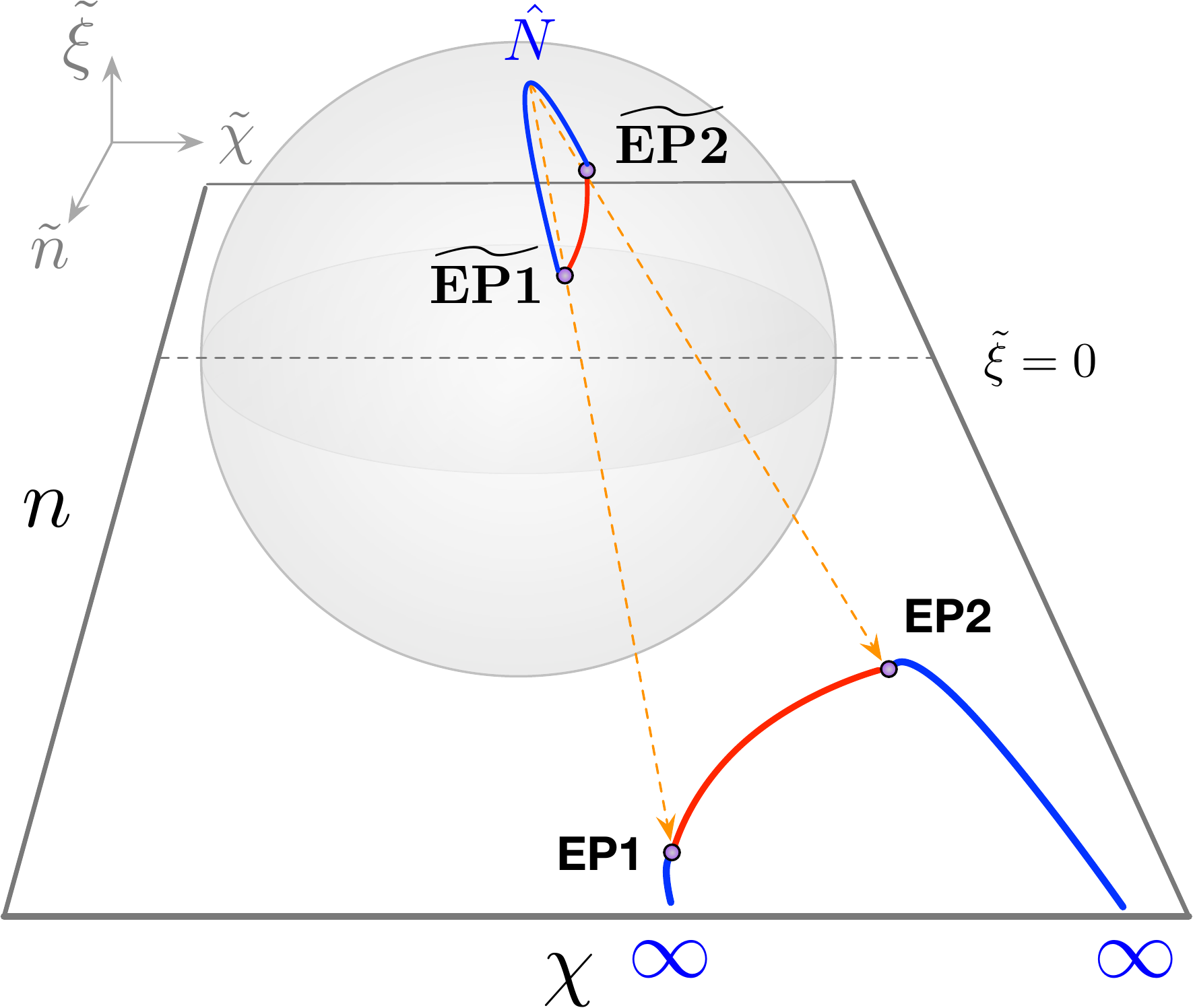}
\caption {A schematic of a Riemann sphere for the branch cuts of the double EPs in an extended parameter plane. The branch cut for the Riemann surfaces in Fig.~\ref{Figure-5}(a) is a curve directly joining two EPs. Two branch cuts for the Riemann surface in Fig.~\ref{Figure-5}(b) meet at infinity.}
\label{Figure-6}
\end{figure}

The Riemann sphere is a mathematical model that describes an extended complex plane. The extended complex plane comprises a complex plane ($\mathbb{C}$) and a point at infinity, which is represented by $\mathcal{C}_{\infty}=\mathbb{C}\cup{\{\infty\}}$. In this section, we exploit this extended complex plane ($\mathcal{C}_{\infty}$) and Riemann sphere to understand the topological structure of double EPs. For this analogy, we consider an isomorphism between the parameter plane $\mathbb{P}=\{(n,\chi)|n, \chi\in\mathbb{R}\}$ and a complex plane $\mathbb{C}$: $\mathbb{P}$ $\cong$ $\mathbb{C}$ defined by ($n,\chi$)$\mapsto$ $n+\iota\chi$. Thus, we can naturally adopt an extended parameter space to utilize a Riemann sphere: $\mathbb{P}_{\infty}$=$\mathbb{P}\cup{\{\infty\}}\cong\mathcal{C}_{\infty}$.

As described in Sec. III, we cannot analytically obtain the eigenvalues ($\lambda_{i}$). However, we can numerically calculate $\lambda_{i}$ as a function of only two parameters ($n,\chi$). Accordingly, we can adopt a functional map ($\widehat{\lambda_{i}}$) from $\mathbb{P}_{\infty}$ to $\mathcal{C}_{\infty}$:
\begin{align}
\widehat{\lambda_{i}}:\mathbb{P}_{\infty}\to \mathcal{C}_{\infty}
\label{eq8}
\end{align}

However, we considered only two specific eigenstates and neglected the other states, but this assumption is only possible within a restricted parameter plane. This is because the two specific eigenstates can interact with other eigenstates in other ranges in the parameter plane. Moreover, infinity cannot be dealt with in a practical physical system. Therefore, we only performed numerical simulations of the restricted range in the parameter plane, as shown in Fig. ~\ref{Figure-5}, where $p=[n_{{\rm in}; 1}, n_{{\rm in}; 2}]\times[\chi_{1}, \chi_{2} ]\subset \mathbb{P}_{\infty}$ with $n_{{\rm in};1}=2.6, n_{{\rm in}; 2}=3.0$ and $\chi_{1}=0.50, \chi_{2}=0.63$, respectively. Thereby, we can define a restricted map $\widehat{\lambda_{i}}|p$: $p \to \mathbb{C}$ with the condition $\lambda_{1}(n,\chi)=\lambda_{2}(n,\chi)$) if and only if $(n,\chi)\in\{\mathbf{EP1}, \mathbf{EP2}\}$.

In order to address the structures of branch cuts that can go to infinity, we should extend the restricted map $\widehat{\lambda_{i}}|p$ to the $\widehat{\lambda_{i}}$ under the certain conditions: (a) $\widehat{\lambda_{i}}((n,\chi)$=$\widehat{\lambda_{i}}|{p}(n,\chi)$ for all $(n,\chi)\in p$; (b) $\widehat{\lambda_{1}}(n,\chi)=\widehat{\lambda_{2}}(n,\chi)$) if and only if $(n,\chi)\in\{\mathbf{EP1}, \mathbf{EP2}\}$; (c) $\widehat{\lambda_{i}}(n=\infty,\chi=\infty)<\infty$. Under these conditions, our $2\times2$ non-Hermitian Hamiltonian is well defined for the double EPs in the extended parameter plane $\mathbb{P}_{\infty}$, and can yield branch cuts at infinity.

We then exploit the Riemann sphere to understand the topological structure of double EPs by stereographic projection. Stereographic projection is a smooth and bijection map that projects a sphere onto a plane. In this study, the Riemann sphere is an embedded manifold in a parameter $\mathbb{R}^{3}$ space that comprises of ($\tilde{n}, \tilde{\chi}, \tilde{\xi}$). Here, $\tilde{\xi}$ is a dummy axis for building up $\mathbb{R}^{3}$ space. Thus, we can consider a unit sphere $\mathbb{S}^{2}$ as a Riemann sphere in the parameter space $\mathbb{R}^{3}$: $\mathbb{S}^{2}=\{(\tilde{n}, \tilde{\chi}, \tilde{\xi})\in \mathbb{R}^{3}$: $\tilde{n}^{2}+\tilde{\chi}^{2}+\tilde{\xi}^{2}=1\}$ and a stereographic projection map $\Pi : \mathbb{S}^{2}\to \mathbb{P}_{\infty}$.

The stereographic projection of $\mathbb{S}^{2}$ is performed from the north pole $\hat{N}=(0, 0, 1)$ onto the plane $\tilde{\xi}=0$ and is explicitly defined as follows:
\begin{align}
\Pi(\tilde{n}, \tilde{\chi}, \tilde{\xi})&=\left(\frac{\tilde{n}}{1-\tilde{\xi}}, \frac{\tilde{\chi}}{1-\tilde{\xi}} \right) \label{eq9} \\
\Pi^{-1}(n,\chi)&=\left(\frac{2n}{\zeta}, \frac{2\chi}{\zeta}, \frac{n^{2}+\chi^{2}-1}{\zeta} \right), \label{eq10}
\end{align}
where $\zeta=n^{2}+\chi^{2}+1$.

 It is found that the double EPs were located at $(n\simeq 2.6257, \chi=0.6001)$ and $(n\simeq 2.9036, \chi=0.5372)$ on $\mathbb{P}_{\infty}$, respectively. Therefore, the double EPs ($\widetilde{\mathbf{EP1}}$ and $\widetilde{\mathbf{EP1}}$) on $\mathbb{S}^{2}$ corresponding to {\bf EP1} and {\bf EP2} on $\mathbb{P}_{\infty}$ are located at $\big(\tilde{n}\simeq 0.7239, \tilde{\chi}\simeq 0.1654, \tilde{\xi}\simeq 0.8621 \big)$ and $\big(\tilde{n}\simeq 0.5974, \tilde{\chi}\simeq 0.1105, \tilde{\xi}\simeq 0.8971\big)$, respectively. Accordingly, we plotted these four points in Fig.~\ref{Figure-6}.

A branch cut is generally a curve that joins two branch points. In our case, the red solid curve on $\mathbb{S}^{2}$ is the branch cut for the Riemann surfaces of $\lambda_{r}$, which directly connects the double EPs, and is represented by a longer curve on $\mathbb{P}_{\infty}$ by $\Pi(\tilde{n}, \tilde{\chi}, \tilde{\xi})$. Furthermore, the other branch cut for the Riemann surfaces of $\lambda_{i}$ denoted by blue solid curve are connected indirectly detour the north pole ($\hat{N}$). Accordingly, its stereographic projection onto $\mathbb{P}_{\infty}$ reveals features that are quite different from the stereographic projection of the Riemann surfaces of $\lambda_{r}$. When approaching the north pole $\hat{N}$, $\tilde{\xi}$ goes to one, making both $n, \chi$ also go to infinity by Eq.~(\ref{eq10}). Note that these two branch cuts on $\mathbb{P}_{\infty}$ meet at infinity, as shown in Fig.~\ref{Figure-6}. Consequently, these two different stereographic projections onto $\mathbb{P}_{\infty}$ explain the two different topological structures of the Riemann sheet in Fig.~\ref{Figure-5}(a) and (b).

\section{CONCLUSIONS}
In this study, we exploited a non-Hermitian coupling of a non-Hermitian Hamiltonian under the convex combination of real and imaginary parts, and analyzed the relationship between the optical microcavity and $2\times2$ toy model. When the real part is dominant, the avoided crossings exhibit Landau--Zener interactions, whereas the imaginary part is dominant, and the avoided crossings exhibit a width bifurcation.

We also investigated the successive transitions of the avoided crossings associated with the width bifurcations in the microcavity, and compared them to the toy model under the variation in the complex coefficient in the range of $0.76\leq\beta\leq 1$ (strong imaginary coupling). Consequently, we identified double EPs on two-level systems in a single microcavity. The topological structures of the Riemann surfaces for these double EPs are equivalent to the form of $f(z)=\{(z-z_{1})(z-z_{2})\}^{1/2}$, and we verified their topological structures by exploiting the Riemann sphere and by encircling the EPs. Their topological structures are significantly different from the typical topological structures of Riemann surfaces in the form of $f(z)=(z-z_{0})^{1/N}$. This is because the former has two singular points at $z=z_{1}$ and $z=z_{2}$, whereas the latter has only one singular point at $z=z_{0}$, even with increasing $N$.

In summary, we implement multiple EPs on two-level systems in a single microcavity by exploiting the strong imaginary coupling of a non-Hermitian Hamiltonian. On the other hand, for the strong real coupling, we require multi-level systems or composite physical systems. Accordingly, our work can aid in advancing the physics associated with EPs and improving performance in applications. More importantly, our mechanism, which employs a strong imaginary coupling of a non-Hermitian Hamiltonian, can be applicable to any kind of non-Hermitian system.

\begin{acknowledgments}
We thank KyeongRo Kim for useful comments. This work was supported by the National Research Foundation of Korea and a grant funded by the Ministry of Science and ICT (Grant No. NRF-2020M3E4A1077861), and Ministry of Education (Grant No. NRF-2021R1I1A1A01052331 \& NRF- 2021R1I1A1A01042199).
\end{acknowledgments}
\bigskip
\section*{Disclosures}
The authors declare no conflicts of interest.



\begin{references}
\bibitem{R09}
I.~Rotter,
``A non-Hermitian Hamilton operator and the physics of open quantum systems,"
J.~Phys.~A: Math.~Theor.~\textbf{42}, 153001 (2009).

\bibitem{W04}
W.~D.~Heiss,
``Exceptional points of non-Hermitian operators,"
J.~Phys.~A: Math.~Gen.~\textbf{37}, 2455 (2004).

\bibitem{T66}
T.~Kato,
Perturbation  Theory  for  Linear  Operators  (Springer, Berlin, 1966).

\bibitem{GL18}
W.~Gao, X.~Li, M.~Bamba, and J.~Kono,
``Continuous transition between weak and ultrastrong coupling through exceptional points in carbon nanotube microcavity exciton--polaritons,"
Nat.~Photon.~\textbf{12}, 362 (2018).

\bibitem{LP18}
L.~M.~de~L\'{e}pinay, B.~Pigeau, B.~Besga, and O.~Arcizet,
``Eigenmode orthogonality breaking and anomalous dynamics in multimode nano-optomechanical systems under non-reciprocal coupling,"
Nat.~Commun.~\textbf{9}, 1401 (2018).

\bibitem{WZ21}
W.~Xiong, Z.~Li, Y.~Song, J.~Chen, G.-Q,~Zhang, and M.~Wang,
``Higher-order exceptional point in a pseudo-Hermitian cavity optomechanical system,"
\pra~\textbf{104}, 063508 (2021).

\bibitem{ZP18}
H.~Zhou, C.~Peng, Y.~Yoon, C.~W.~Hsu, K.~A.~Nelson, L.~Fu, J.~D.~Joannopoulos, M.~Solja\u{c}i\'{c}, and B.~Zhen,
``Observation of bulk Fermi arc and polarization half charge from paired exceptional points,"
Science~\textbf{359}, 1009 (2018).

\bibitem{CY21}
C.~F.~Fong, Y.~Ota, Y.~Arakawa, S.~Iwamoto, and Y.~K.~Kato,
``Chiral modes near exceptional points in symmetry broken H1 photonic crystal cavities,"
Phys.~Rev.~Res.~\textbf{3}, 043096  (2021).

\bibitem{WS17}
W.~Chen, \c{S}.~K.~\"{O}zdemir, G.~Zhao, J.~Wiersig, and L.~Yang,
``Exceptional points enhance sensing in an optical microcavity,"
Nature~\textbf{548}, 192 (2017).

\bibitem{CW21}
C.~Wang, W.~R.~Sweeney, A.~D.~Stone, and L.~Yang,
``Coherent perfect absorption at an exceptional point,"
Science~\textbf{373}, 1261 (2021).

\bibitem{ZL17}
D.~Zhang, X.-Q.~Luo, Y.-P.~Wang, T.-F.~Li, and J.~Q.~You,
``Observation of the exceptional point in cavity magnon-polaritons,"
Nat. Commun.~\textbf{8}, 1368 (2017).

\bibitem{HD19}
H.~Liu, D.~Sun, C.~Zhang, M.~Groesbeck, R.~Mclaughlin, and Z.~V.~Vardeny,
``Observation of exceptional points in magnonic parity-time symmetry devices,"
Sci. Adv.~\textbf{5}, eaax9144 (2019).

\bibitem{CH18}
Y.~Choi, C.~Hahn, J.~W.~Yoon, and S.~H.~Song,
``Observation of an anti-PT-symmetric exceptional point and energy-difference conserving dynamics in electrical circuit resonators,"
Nat. Commun.~\textbf{9}, 2182 (2018).

\bibitem{RK18}
R.~El-Ganainy, K.~G.~Makris, M.~Khajavikhan, Z.~H.~Musslimani, S.~Rotter, and D.~N.~Christodoulides,
``Non-Hermitian physics and PT symmetry,"
Nat.~Phys.~\textbf{14}, 11 (2018).

\bibitem{LT21}
L.~Xiao, T.~Deng, K.~Wang, Z.~Wang, W.~Yi, and P.~Xue
``Observation of Non-Bloch Parity-Time Symmetry and Exceptional Points,"
\prl~\textbf{126}, 230402  (2021).

\bibitem{MY16}
M.~I.~Afzal and Y.~T.~Lee,
``Supersymmetrical bounding of asymmetric states and quantum phase transitions by anti-crossing of symmetric states,"
Sci.~Rep.~\textbf{6}, 39016 (2016).

\bibitem{AR21}
A.~Bergman, R.~Duggan, K.~Sharma, M.~Tur, A.~Zadok, and A.~Al\`{u}
``Observation of anti-parity-time-symmetry, phase transitions and exceptional points in an optical fibre,"
Nat.~Commun.~\textbf{12}, 486 (2021).

\bibitem{TG18}
T.~Gao, G.~Li, E.~Estrecho, T.~C.~H.~Liew, D.~Comber-Todd, A.~Nalitov, M.~Steger, K.~West, L.~Pfeiffer, D.~W.~Snoke, A.~V.~Kavokin, A~G.~Truscott, and E.~A.~Ostrovskaya,
``Chiral Modes at Exceptional Points in Exciton-Polariton Quantum Fluids,"
\prl~\textbf{120}, 065301 (2018).

\bibitem{HG22}
H.~Nasari, G.~Lopez-Galmiche, H.~E.~Lopez-Aviles, A.~Schumer, A.~U~Hassan, Q.~Zhong, S.~Rotter, P.~LiKamWa, D.~N.~Christodoulides, and M.~Khajavikhan,
``Observation of chiral state transfer without encircling an exceptional point,"
Nature~\textbf{605}, 256 (2022).

\bibitem{KJ18}
K.-W.~Park, J.~Kim, S.~Moon, and K.~An,
``Maximal Shannon entropy in the vicinity of an exceptional point in an open microcavity,"
Sci.~Rep.~\textbf{10}, 12551 (2020).

\bibitem{J14}
J.~Wiersig,
``Enhancing the Sensitivity of Frequency and Energy Splitting Detection by Using Exceptional Points: Application to Microcavity Sensors for Single-Particle Detection,"
\prl~\textbf{112}, 203901 (2014).

\bibitem{RJ22}
R.~Kononchuk, J.~Cai, F.~Ellis, R.~Thevamaran, and T.~Kottos,
``Exceptional-point-based accelerometers with enhanced signal-to-noise ratio,"
Nature~\textbf{607}, 697 (2022).

\bibitem{MM18}
M.~J.~Grant and M.~J.~F.~Digonnet,
``Rotation sensitivity and shot-noise-limited detection in an exceptional-point coupled-ring gyroscope,"
Opt.~Lett.~\textbf{46}, 2936 (2021).

\bibitem{MA19}
M.~P.~Hokmabadi, A.~Schumer, D.~N.~Christodoulides, and M.~Khajavikhan,
``Non-Hermitian ring laser gyroscopes with enhanced Sagnac sensitivity,"
Nature~\textbf{576}, 70 (2019).

\bibitem{DM16}
J.~Doppler, A.~A.~Mailybaev, J.~B\"{o}hm, U.~Kuhl, A.~Girschik, F.~Libisch, T.~J.~Milburn, P.~Rabl, N.~Moiseyev, and S.~Rotter,
``Dynamically encircling an exceptional point for asymmetric mode switching,"
Nature~\textbf{537}, 76 (2016).

\bibitem{XA22}
X.~Shu, A.~Li, G.~Hu, J.~Wang, A.~Al\`{u}, and L.~Chen,
``Fast encirclement of an exceptional point for highly efficient and compact chiral mode converters,"
Nat.~Commun.~\textbf{13}, 2123 (2022).

\bibitem{HA17}
H.~Hodaei, A.~U.~Hassan, S.~Wittek, H.~Garcia-Gracia, R.~El-ganainy,  D.~N.~Christodoulides, and M.~Khajavikhan,
``Enhanced sensitivity at higher-order exceptional points,"
Nature~\textbf{548}, 187 (2017).

\bibitem{IE22}
I.~Mandal and E.~J.~Bergholtz,
``Symmetry and Higher-Order Exceptional Points,"
\prl~\textbf{127}, 186601 (2021).

\bibitem{HG16}
K.~Ding, G.~Ma, M.~Xiao, Z.~Q.~Zhang, and C.~T.~Chan,
``Emergence, Coalescence, and Topological Properties of Multiple Exceptional Points and Their Experimental Realization,"
Phys.~Rev.~X~\textbf{6}, 021007 (2016).

\bibitem{SA20}
S.~Dey, A.~Laha, and S.~Ghosh,
``Nonlinearity-induced anomalous mode collapse and nonchiral asymmetric mode switching around multiple exceptional points,"
\prb~\textbf{101}, 125432 (2020).

\bibitem{JS12}
S.-Y.~Lee, J.-W.~Ryu, and S.~W.~Kim,
``Analysis of multiple exceptional points related to three interacting eigenmodes in a non-Hermitian Hamiltonian,"
\pra~\textbf{85}, 042101 (2012).

\bibitem{FZ21}
F.~Yu, X.-L.~Zhang, Z.-N.~Tian, Q.-D.~Chen, and H.-B.~Sun,
``General Rules Governing the Dynamical Encircling of an Arbitrary Number of Exceptional Points,"
\prl~\textbf{127}, 253901 (2021).

\bibitem{SJ12}
S.-Y.~Lee, J.-W.~Ryu, S.~W.~Kim, and Y.~Chung,
``Geometric phase around multiple exceptional points,"
\pra~\textbf{85}, 064103 (2012).

\bibitem{RD22}
R.~Nehra and D.~Roy,
``Topology of multipartite non-Hermitian one-dimensional systems,"
\prb~\textbf{105}, 195407 (2022).

\bibitem{XC19}
X.-L. Zhang and C.~T.~Chan,
``Dynamically encircling exceptional points in a three-mode waveguide system,"
Commun.~Phys.~\textbf{2}, 63 (2019).

\bibitem{AK22}
A.~Hashemi, K.~Busch, D.~N.~Christodoulides, S.~K.~Ozdemir, and R.~El-Ganainy,
``Linear response theory of open systems with exceptional points,"
Nat.~Commun.~\textbf{13}, 3281 (2022).

\bibitem{S08}
S.-Y. Lee, J.-W. Ryu, J.-B. Shim, S.-B. Lee, S. W. Kim, and K. An,
``Divergent Petermann factor of interacting resonances in a stadium-shaped microcavity,"
\pra~\textbf{78}, 015805 (2008).

\bibitem{WA90}
W.~Heiss and A.~Sannino,
``Avoided level crossing and exceptional points,"
J.~Phys.~A: Math.~Gen.~\textbf{23}, 1167 (1990).

\bibitem{HI13}
H.~Eleuch and I.~Rotter,
``Width bifurcation and dynamical phase transitions in open quantum systems,"
\pre~\textbf{87}, 052136 (2013).

\bibitem{HI14}
H.~Eleuch and I.~Rotter,
``Open quantum systems and Dicke superradiance,"
Eur. Phys. J. D~\textbf{68}, 74 (2014).

\bibitem{W03}
J.~Wiersig,
``Boundary element method for resonances in dielectric microcavities,"
J.~Opt.~A: Pure~Appl.~Opt.~\textbf{5}, 53 (2003).

\bibitem{H99}
H.-J.~St\"{o}ckmann,
Quantum Chaos: An Introduction (Cambridge University Press, London, 1999).

\bibitem{F10}
F.~Haake,
Quantum Signatures of Chaos (Springer, Berlin, 2010).

\bibitem{JE29}
J.~von Neumann and E.~Wigner,
Phys.~Z.~\textbf{30}, 467 (1929).

\bibitem{KS18}
K.-W.~Park, S.~Moon, Y.~Shin, J.~Kim, K.~Jeong, and K.~An,
``Shannon entropy and avoided crossings in closed and open quantum billiards,"
\pre~\textbf{97}, 062205 (2018).

\bibitem{IJ15}
I.~Rotter and J.~P.~Bird,
``A review of progress in the physics of open quantum systems: theory and experiment,"
Rep.~Prog.~Phys.~\textbf{78} 114001 (2015).

\bibitem{B15}
B. Zhen, C. W. Hsu, Y. Igarashi, L. Lu, I. Kaminer, A. Pick, S.-L. Chua, J. D. Joannopoulos, and M. Solja\v{c}i\'{c},
``Spawning rings of exceptional points out of Dirac cones,"
Nature~\textbf{525}, 354 (2015).


\end{references}
\end{document}